# Breaking Peierls' theorem in polyacetylene chains *via* topological design

Xinnan Peng[1,10], Marco Lozano[2,10], Jie Su[3,10], Lulu Wang[1,10], Diego Soler-Polo[2], Thomas Tuloup[4], Junting Wang[5], Shaotang Song[1], Ming Wah Wong[1], Jiangbin Gong[4,6], Junzhi Liu[5], Franz J Giessibl[7], Pavel Jelínek[2,8]*, Jiong Lu[1,9]*

[1]Department of Chemistry, National University of Singapore, 3 Science Drive 3, Singapore 117543, Singapore.

[2]Institute of Physics of the Czech Academy of Science, 16200 Praha, Czech Republic.

[3]Department of Applied Physics, The Hong Kong Polytechnic University, Hung Hom, Kowloon, Hong Kong, People's Republic of China.

[4]Department of Physics, National University of Singapore, 2 Science Drive 3, Singapore 117551, Singapore.

[5]Department of Chemistry and State Key Laboratory of Synthetic Chemistry, The University of Hong Kong, 999077 Hong Kong, People's Republic of China.

[6]Centre for Quantum Technologies, National University of Singapore, 3 Science Drive 2, Singapore 117543, Singapore.

[7]Institute of Experimental and Applied Physics, University of Regensburg, 93053 Regensburg, Germany.

[8]Regional Centre of Advanced Technologies and Materials, Palacký University, 78371 Olomouc, Czech Republic.

[9]Institute for Functional Intelligent Materials, National University of Singapore, 4 Science Drive 2, Singapore 117544, Singapore.

[10]These authors contributed equally: Xinnan Peng, Marco Aurelio Lozano, Jie Su, Lulu Wang.

E-mail: jelinekp@fzu.cz (P.J.); chmluj@nus.edu.sg (J.L.)

**Abstract**:

Peierls theorem postulates that a one-dimensional (1D) metallic chain must undergo a metal-to-insulator transition *via* lattice distortion, resulting in bond length alternation (BLA) within the chain[1]. The validity of this theorem has been repeatedly proven in practice, as evidenced by the absence of a metallic phase in low-dimensional atomic lattices and electronic crystals, including π-conjugated polymers[2,3], artificial 1D quantum nanowires[4-6], and anisotropic inorganic crystals[7,8]. Overcoming this transition enables realizing long-sought organic quantum phases of matter, including 1D synthetic organic metals and even high-temperature organic superconductors[9-11]. Herein, we demonstrate that the Peierls transition can be globally suppressed by employing lattice topology engineering of classic trans-polyacetylene chains connected to open-shell nanographene terminals. The appropriate topology connection enables an effective interplay between the zero-energy modes (ZMs) of terminal and the finite odd-membered polyacetylene (OPA) chains. This creates a critical topology-defined highest occupied molecular orbital (HOMO) that compensates for bond density variations, thereby suppressing BLA and reestablishing their quasi-1D metallic character. Moreover, it also causes the formation of an unconventional *boundary-free resonance state*, being delocalized over the entire chain with non-decaying spectral weight, distinguishing them from traditional solitons observed in polyacetylene. Our finding sets the stage for pioneering the suppression of material instability and the creation of synthetic organic quantum materials with unconventional quantum phases previously prohibited by the Peierls transition.

**Introduction**

Understanding the structural and electronic instabilities of matter holds significant importance in condensed matter physics, chemistry, and material science, as these instabilities often lead to symmetry breaking transitions and alterations in material properties. According to the well-established Peierls' theorem[1], equidistant 1D metallic chains are energetically unstable because of the divergence of response function due to the perfect nesting of the Fermi surface[12-14]. As a result, below the critical temperature, these 1D chains undergo a metal-to-insulator transition, accompanied by periodic lattice distortion known as bond length alternation (BLA). Above the critical temperature, the Kohn anomaly at phonon spectra will also emerge[15]. The energy penalty associated with the lattice distortions is offset by the electronic energy gained from the band gap opening, thus stabilizing the broken-symmetry solution with the characteristic BLA structure[1]. Circumvention of these instabilities by suppressing the Peierls transition would unlock exciting opportunities for designing metallic states and exotic quantum phases[9-11], not only in 1D systems but also extendable to higher dimensions[16].

To date, it has only been postulated that the Peierls BLA can be suppressed locally near a domain wall (commonly referred to as a soliton) in odd-membered trans-polyacetylene (OPA) chains (Fig. 1a). However, BLA persists throughout the rest of the OPA chain, which exhibits doubly degenerate ground states (Fig. 1e). An OPA chain hosts a soliton with one zero-energy mode (ZM), which can be viewed as a spatially localized domain wall separating two distinct phases with mirror symmetry and identical system energies (Fig. 1a)[17-19]. In an infinite chain, the position of the domain wall (soliton) does not affect the overall system energy, allowing it to propagate freely along the chain in response to external excitations[8,18]. However, in finite chains, the soliton tends to localize in the central region (Fig. 1c), encompassing only a few adjacent units. In contrast, if the BLA can be sufficiently suppressed over the entire chain (Fig. 1b), it would result in 1D metallicity with a metallic band crossing the Fermi level ($E_F$) and an unconventional *boundary-free resonance state* (termed as *BFRS*, Fig. 1d) exhibiting non-decaying spectral weight at each site. Despite significant efforts in recent decades towards the synthesis of OPA chains and 1D polymers hosting solitons[20-23], the fundamental design principles for such 1D systems, which violate Peierls' theorem, remain unexplored. Consequently, the experimental realization of these 1D systems with fully suppressed BLA has yet to be achieved.

The emergence of the Peierls transition could be rationalized by the tight-binding 1D Peierls model (see Supplementary Information Section 1.1 and 1.2)[24]

$$\mathcal{H}_{Peierls} = -2 \sum_n t(\delta_n) \hat{T}_n + \mathcal{H}_{elastic} \quad (1),$$

where the first term represents the electronic kinetic energy and the second term is the elastic energy $\mathcal{H}_{elastic}$ associated with the deformation $\delta_n$ of individual bonds in the chains. Here $t$ means the electronic hopping between neighbor atomic sites of $n$-th bond, which is function of the bond deformation $\delta_n$. $\hat{T}_n = 1/2 \, (c_{n+1}^+ c_n + c_n^+ c_{n+1})$ represents the bond density operator, whose expectation value $\langle \hat{T}_n \rangle$ determines the electronic bond density accumulated in $n$-th bond. Importantly, the self-consistent solution of eq. (1), which is discussed in detail in Supplementary Information, reveals that the deformation $\delta_n$ of individual bonds is directly proportional to the corresponding bond densities $\langle \hat{T}_n \rangle$, reflecting the (anti)bonding character of occupied molecular orbitals on a particular $n$-th bond. Consequently, at the ground state, each bond has different bond densities, resulting in different bond lengths across the chain. In particular, bonds with higher bond density become shorter (double bonds), while those with lower bond density become longer (single bonds), as shown in Fig. 1a. Thus, the Peierls transition can be directly linked to the variation of bond densities $\langle \hat{T}_n \rangle$ within the chain, which destabilizes the equidistant bond lengths and causes the BLA order. To prevent the Peierls transition, the key is to minimize variation in bond densities across the chain (Fig. 1b).

In recent years, topological phases of matter have emerged as a central concept in condensed matter physics, leading to the discovery of topological insulators with protected in-gap states and topological superconductors[3,25-28]. Depending on the winding number, a topologically non-trivial insulating phase can exist in finite 1D chains, enabling the formation of protected end states against external perturbation. Additionally, π-magnetism of nanographenes can be anticipated from the emergence of ZM states, a consequence of the topological properties of the honeycomb bipartite lattice[29-33]. The following discussion demonstrates that the interaction between the topological ZMs of nanographenes and the OPA chain enables the suppression of the classic Peierls transition, thereby realizing unconventional quantum quasi-metallic states in 1D chains[34].

**The design principles**

Here, we introduce a design strategy involving the lattice topology engineering of OPA chains to defeat the Peierls transition, achieving a quasi-metallic state and a *boundary-free resonance state* (*BFRS*) delocalized across the entire structure. This approach consists of terminating the OPA chain by nanographenes with a matched topology of their bipartite lattice, which gives rise to ZMs. The 1D Peierls model combined with DFT calculations of multiple examples enables us to establish simple rules leading to the violation of the Peierls transition in OPA chains symmetrically connected to both open-shell benzenoid and non-benzenoid nanographenes (BNGs and NBNGs), whereby bond density variations in the OPA chains can be completely suppressed. Specifically, we demonstrate that an appropriate connection leads to the interaction between the ZMs of nanographenes and the ZM of OPA chain (Extended Data Fig. 1), resulting in the formation of a new electronic highest occupied molecular orbital (HOMO), which compensates for the bond density variations in the OPA chain. Consequently, this results in the suppression of the Peierls transition along with the formation of a *BFRS*. Moreover, for sufficiently long OPA chains, we observe theoretically the emergence of a metallic band across the Fermi level.

We first illustrate this design principle in a seven-carbon atom OPA chain with both ends connected with a [2]triangulene molecule[35], a simple, representative open-shell nanographene that features one ZM due to sublattice imbalance (Fig. S2 and Supplementary Information Section 1.3). Fig. 2 shows two possible connections of [2]triangulene at the chain ends: **Peierls-Tri-[7]PA** (connecting at the minor sublattice without ZM density of [2]triangulene, Fig. 2a) and **Tri-[7]PA** (connecting at the major sublattice with ZM density of [2]triangulene, Fig. 2b). Our theoretical analysis reveals that **Peierls-Tri-[7]PA** shows significant bond density variation and BLA, indicating the presence of the Peierls transition (Fig. 2a). In contrast, **Tri-[7]PA** exhibits the absence of bond density variation and BLA, indicating the suppression of the Peierls transition (Fig. 2b). Despite their isomeric nature, **Peierls-Tri-[7]PA** and **Tri-[7]PA** exhibit different sublattice imbalances $|N_A - N_B|$. Notably, **Peierls-Tri-[7]PA** hosts three degenerate ZMs in the Hückel energy spectrum, while **Tri-[7]PA** features only one ZM (Fig. 2c-e), in accordance with the mirror theorem[29]. Both structures show nearly identical low-energy occupied electronic states (Fig. S3), except for their frontier orbitals. Importantly, a new HOMO (Fig. 2e) emerges from the hybridization of original ZMs in **Tri-[7]PA** (Extended Data Fig. 1), but it is absent in **Peierls-Tri-[7]PA**. Notably, the presence of an OPA chain as the central segment is essential, as an even-membered chain prevents effective ZM hybridization (see Supplementary Information Section 1.3 and Fig.

S4). To understand the crucial role of this new HOMO in diminishing bond density variation in the OPA chain, two key points must be noted. First, ZMs are localized exclusively on one sublattice and, therefore, do not contribute to bond density. Second, all occupied molecular orbitals that contribute to the bond densities in **Peierls-Tri-[7]PA** and **Tri-[7]PA** are identical, except for the new HOMO (Fig. S5). Therefore, the contribution of this HOMO in **Tri-[7]PA** is directly responsible for the suppression of the Peierls transition by compensating for the bond density variation. To further reinforce our conclusions, we performed many-body Hubbard model calculations by constructing a Hubbard Hamiltonian, which confirms the robustness of our model (Supplementary Information Section 1.6).

We also observed similar behaviors for other open-shell benzenoid nanographene terminals with specific lattice topologies, including [3]triangulene with two ZMs[36], as well as Clar's goblet[37], which has two ZMs due to the topological frustration of the bipartite lattice (Fig. S6-8 and Supplementary Information Section 1.3). Additionally, this effect is independent of the length of OPA chains. Both Peierls model and DFT calculations reveal a complete suppression of the Peierls transition in an extended OPA chain consisting of 23 carbon atoms or beyond (e.g. 51 carbon atoms) only when the lattice topology matches that of **Tri-[7]PA** (Fig. 2g and Extended Data Fig. 2). Otherwise, only a few bond lengths in the middle are identical, while BLA appears along the rest of the chain. These results align with the distinct distributions of solitons and *BFRS* influenced by the lattice topology. As shown in Extended Data Fig. 3, both the soliton states in a pure OPA chain (51 carbon atoms) without terminals and in **Peierls-Tri-[51]PA**, exhibit their highest probabilities at the midpoint, gradually decaying towards the ends, as expected for conventional solitons (Fig. 1c)[18]. In contrast, **Tri-[51]PA** exhibits a uniform distribution of the resonance state at each site along the entire OPA chain, aligning with the *BFRS* picture (Fig. 1d). This finding suggests that in-gap states manifest either localization or resonance depending on the lattice topology. In addition, the delocalization of the *BFRS* is independent of the band gap, unlike conventional solitons, whose delocalization decreases as the band gap increases[38]. Furthermore, we observe that as the length of the OPA chain with topology-matched terminals increases, the number of electronic states near the $E_F$ also increases (Extended Data Fig. 4d-f). In principle, this leads to a continuous density of states (DOS) across the $E_F$ for sufficiently long polyacetylene chains, revealing the metallic character. In contrast, OPA chains lacking the appropriate lattice topology consistently exhibit a band gap (Fig. 2f and Extended Data Fig. 4a-c). Thus, we demonstrate the possibility of constructing a purely 1D metallic state in polyacetylene-based systems through lattice topology engineering.

Therefore, to suppress the Peierls transition and achieve unconventional metallic phase in OPA chains, the following criteria must be met: (1) topology-matched BNG terminal must host at least one singly occupied ZM, arising from sublattice imbalance or topological frustration of the bipartite lattice; (2) the terminals must be connected to the OPA chain through sites where their ZMs are present, reducing the total number of ZMs in the system by two compared to the combined ZMs of the individual BNGs and the OPA chain; (3) the hybridization of ZMs must generate a new HOMO that compensates for bond density variation, thereby suppressing the Peierls transition in the OPA chain. Furthermore, this general design principle can be extended to open-shell terminals comprising non-benzenoid nanographenes (NBNGs), as discussed later.

**On-surface synthesis of nanographene terminated OPA chains**

Inspired by these predictions, we explored the synthesis of a series of nanographene-terminated OPA chains of varying lengths. The molecular building block for fabricating these OPA chains was synthesized through on-surface synthesis method as reported in our previous study. Specifically, a nanographene terminated by a trans-C-C chain with alternating single-double bonds can be fabricated on Au(111) through a highly selective ring-opening of the seven-membered ring within the azulene moieties[39]. PA chains with symmetric nanographene terminals can form through intermolecular dimerization between two C-C chains of the building blocks (Extended Data Fig. 5a). After further thermal annealing at 500 K, subsequent STM imaging revealed that the Au(111) surface was decorated with dumbbell-shaped molecular species, along with various isomers exhibiting different STM contrasts (Extended Data Fig. 5b-h). It is worth noting that thermally induced detachment of carbon atoms leads to the formation of C-C chains of varying lengths including a series of OPA chains. Non-contact atomic force microscopy (NC-AFM) imaging with a CO-functionalized tip reveals the C-C chain made of seven carbon atoms with nanographene terminals consisting of 5-, 6-, and 7-membered rings (Fig. 3e). The perfect zigzag-shaped topology of the C-C chain, together with the absence of any additional protrusions or nodes, signifies its trans-polyacetylene chain structure. Furthermore, the bright features observed on the pentagons in terminal nanographene indicate the $CH_2$ termination[36,39-41], while the molecular backbone remains planar upon adsorption on Au(111) (Fig. S17). The chemical structures of these OPA chains identified from the AFM images align well with their simulated AFM images (Fig. S18)[42,43], confirming the successful synthesis of three-membered, five-membered, seven-membered, PA chains with nanographene terminals, referred to as **[3]PA**, **[5]PA**, and **[7]PA**, respectively.

**Suppression of the Peierls transition in synthesized OPA chains**

Our calculations indicate that the original structure of the terminal in **[7]PA** hosts a singly occupied molecular orbital (SOMO), reminiscent of ZM in benzenoid structures (Fig. S10).

The synthesized **[7]PA** features an OPA chain symmetrically terminated by open-shell nanographenes, allowing potential suppression of the Peierls transition depending on the end topology, as discussed above. To examine the Peierls transition in **[7]PA**, we performed a series of meticulous bond length measurements using NC-AFM, which can qualitatively differentiate the bond lengths in similar local environments by probing the electron density[44-46]. As shown in Fig. 3m,q, the averaged apparent bond length of the C-C bonds in the OPA is 1.30 Å ± 0.06 Å, with no discernible variation of bond lengths evident in the Laplace-filtered AFM image. In addition, the DFT-calculated bond lengths are uniform at 1.39 Å (Fig. 3i), in accordance with the measured values. These results indicate the absence of BLA in the OPA due to the effective suppression of the Peierls transition.

Furthermore, as shown in Fig. 3b,f, tip-induced dehydrogenation of the $sp^3$ carbon in the pentagon of **[7]PA** can be verified by the absence of a bright feature on the top pentagon in the AFM image of the product after tip-manipulation (referred to as **[7]PA-1H**)[36,40]. This enables the breaking of structural symmetry and alteration of the lattice topology, effectively leading to a resurgence of the Peierls transition and BLA in the resulting product, as revealed in the corresponding Laplace-filtered AFM image (Fig. 3n, with indicated bond indexes). It is noted that the first C-C bond (indexed as 1) in the OPA chain connected to the nanographene without a $sp^3$ carbon is notably longer than its neighboring C-C bond (indexed as 2). As shown in Fig. 3r, the C-C bonds with odd index numbers clearly display longer apparent bond lengths, averaging at 1.50 Å ± 0.06 Å as measured by NC-AFM. In contrast, the even-indexed bonds have a shorter apparent average bond length of 1.05 Å ± 0.05 Å. Although it has been reported that CO bending can potentially cause apparent bond distortions, our BLA analysis is based on the relative differences between apparent bond lengths, rather than their absolute values[44,45,47]. These results indicate that the odd-indexed bonds exhibit a localized single bond nature, while even-indexed bonds display a localized double bond nature. Furthermore, the DFT calculations for **[7]PA-1H** (Fig. 3j) also reveal alternated bond lengths of 1.41-1.42 Å (single bond) and 1.36-1.37 Å (double bond) within the OPA chain.

In addition, performing the same bond length analysis for shorter central OPA chains also confirms the absence of BLA in both **[3]PA** and **[5]PA**, connected with the same nanographene terminal (Fig. 3 and Extended Data Fig. 6). Capturing more uniform bond length distribution in **[3]/[5]/[7]PA** suggests the successful suppression of the Peierls transition in their OPA chains with a specific end topology. This is in stark contrast with the noticeable BLA in dehydrogenated structures (**[3]/[5]/[7]PA-1H**) captured by NC-AFM imaging and DFT calculations. We then performed further measurements on **[3]/[5]/[7]PA-2H**, the dehydrogenation products by removing the additional hydrogen atoms on both terminals. This modified terminal lacks a ZM for hybridization. Therefore, a conventional soliton and the Peierls transition are expected in **[3]/[5]/[7]PA-2H**. However, high-resolution AFM imaging of **[3]/[5]/[7]PA-2H** could not be achieved due to the instability (Fig. S19). Furthermore, the Peierls transition persists in the even-membered polyacetylene chains, as confirmed by the observed BLA (Extended Data Fig. 7). This further underscores the essential role of ZM hybridization determined by the lattice topology, as discussed above.

**Characterizations of *boundary-free resonance states* (*BFRS*)**

We then performed differential conductance spectroscopy (d$I$/d$V$) measurements of these finite OPA chains to investigate their electronic properties. Fig. 4a presents the d$I$/d$V$ spectra collected over the central OPA chains of **[3]PA** (red curve), **[5]PA** (blue curve), and **[7]PA** (green curve), alongside the reference spectrum recorded on bare Au(111) (gray curve). Apart from the frontier HOMO and LUMO, all d$I$/d$V$ spectra exhibit a highly delocalized state near $E_F$, tentatively assigned as the *BFRS*. Unlike the Kondo resonance characterized by a sharp zero-bias peak feature, the near-$E_F$ states observed for these OPA chains exhibit much broader peaks centered around 20 meV (**[3]PA**), 40 meV (**[5]PA**), and 40 meV (**[7]PA**), respectively. The absence of Kondo screening indicates the quenching of the radical character of the *BFRS*, which can be attributed to the emergence of the mixed-valence regime due to Fermi-level pinning[38]. The d$I$/d$V$ maps acquired at the corresponding energies reveal the characteristic lobe patterns predominantly localized on the central OPA chains (Fig. 4b-d), as well as their less pronounced local density of states (LDOS) intensity distributed over the nanographene terminals, which align well with the simulated patterns of *BFRS* (Fig. 4e-g, see detailed DFT calculations in Fig. S14). Furthermore, contour plot of d$I$/d$V$ spectra acquired across the molecules reveal nearly constant intensity along the OPA sites (Extended Data Fig. 8j-l), consistent with the expected non-decaying electronic spectral weight of a *BFRS*. As expected, dehydrogenated molecules (**[3]/[5]/[7]PA-1H**) with the restored Peierls transition do not reveal such a metallic-like state near $E_F$ (Fig. 4h), consistent with the DFT calculations (Fig. S15). The corresponding d$I$/d$V$ maps of HOMO and LUMO for both pristine molecules and their dehydrogenated products are presented in Extended Data Fig. 9. Additionally, d$I$/d$V$ measurements of the even-membered polyacetylene chains also confirm the absence of any in-gap states (Extended Data Fig. 10).

**Theoretical rationalization of the suppression of the Peierls transition**

As previously discussed, the complete suppression of the Peierls transition and the construction of *BFRS* in OPA chains can be achieved by the appropriate topological design of benzenoid nanographene terminals. To rationalize the experimental observations presented above, we qualitatively extended our model to non-benzenoid systems lacking bipartite symmetry (see Supplementary Information Section 1.4). As shown in Fig. 5b and Fig. S11, the model Hamiltonian calculations indicate that the bond lengths of all C-C bonds in the OPA chain in **[7]PA** are uniformly 1.39 Å, aligning with NC-AFM measurements and DFT calculations (Fig. 3). In contrast, modifying the end topology by changing the connection site to neighboring carbon (site B, as defined in Fig. S10) with a weaker density of pseudo-ZM, yields another structure exhibiting alternating bond lengths (referred to as **Peierls-[7]PA**). Moreover, the HOMO of **[7]PA** (Fig. 5e) exhibits a similar orbital pattern to the aforementioned unique HOMO of **Tri-[7]PA**, also compensating for the bond density variation (Fig. S12). The analogous electronic structures of **[7]PA** and **Tri-[7]PA** indicate that they share the same class of lattice topology, which is crucial for suppressing the Peierls transition. Notably, the unique HOMO disappears when the end topology is altered to that of **Peierls-[7]PA** (Fig. 5c). As discussed in the benzenoid cases, the lattice topology can influence sublattice imbalances, leading to different numbers of ZMs and adjusting degeneracy of frontier orbitals. Despite the absence of any ZM in non-benzenoid structures due to the lack of bipartite symmetry, engineering the lattice topology can still influence the hybridization of pseudo-ZMs, leading to different numbers of nearly degenerated states. Specifically, the three states (HOMO/SOMO/LUMO) of **Peierls-[7]PA** are much closer in energy position compared to those of **[7]PA** (Fig. 5d and Fig. S11). In addition, these states of **Peierls-[7]PA** resemble the singly-occupied ZM, with electron density localized on every second carbon atom, if considering the central OPA chain with bipartite symmetry (Fig. 5c). These findings reveal that the HOMO, SOMO, and LUMO of **Peierls-[7]PA** are nearly degenerated, in contrast to the non-degenerate frontier orbitals of **[7]PA**. This indicates the different numbers of pseudo-ZMs of **Peierls-[7]PA** (three) and **[7]PA** (one) depending on the end topology. Therefore, we identify a strong analogy with the benzenoid cases, where the suppression of the Peierls transition can be achieved through two key mechanisms: (1) reducing the degeneracy of frontier orbitals *via* lattice topology engineering and (2) the emergence of a unique HOMO orbital, which can vanish completely bond density variation and BLA.

**Conclusion**

Our study presents a novel design principle involving lattice topology engineering of OPA chains to suppress completely the Peierls transition. By introducing a suitable end connection, the effective interaction between the ZMs of terminals and the ZM of finite OPA chain leads to a critical HOMO state, which compensates for bond density variation, thereby suppressing BLA. This approach enables the creation of quasi-metallic 1D chains hosting *boundary-free resonance states* with non-decaying electronic character that are delocalized across the entire structure. The lattice topology engineering demonstrated here opens new avenues for stabilizing unconventional 1D quantum phases of matter, paving the way for the fabrication of synthetic organic metals and superconductors.

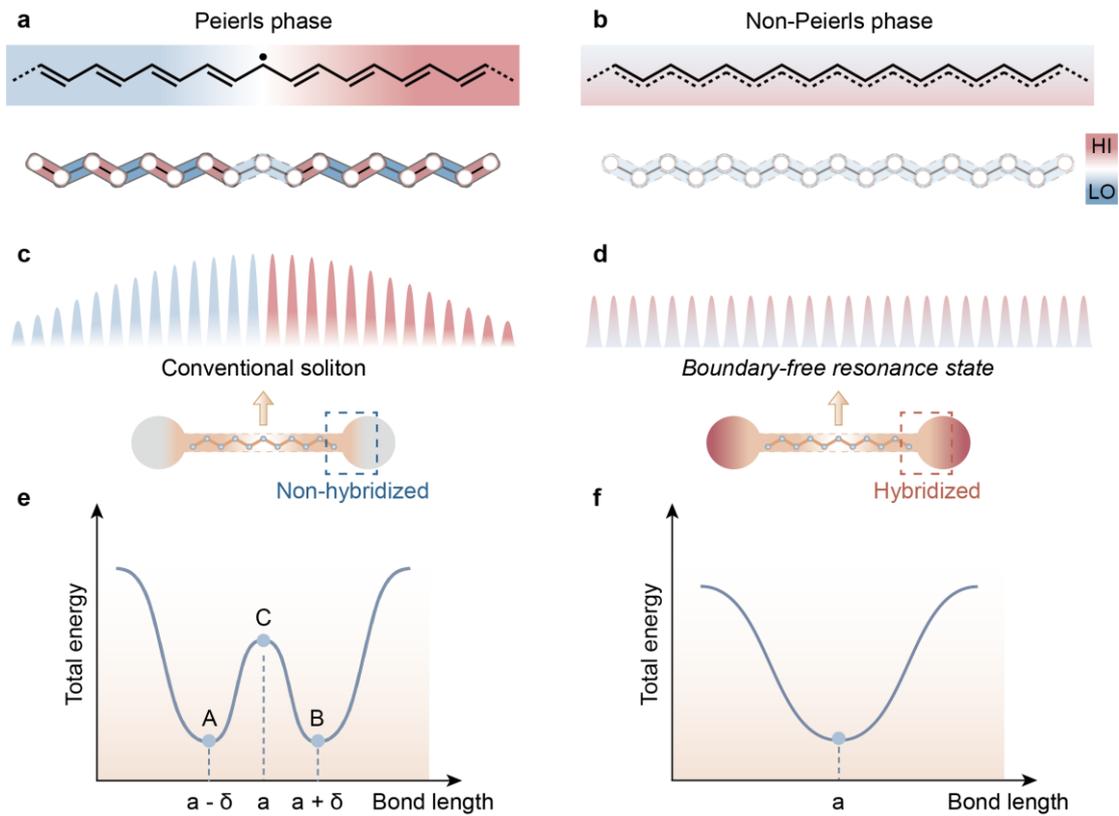

**Fig. 1 | Suppression of the Peierls transition in OPA chains and emergence of a *boundary-free resonance state* (*BFRS*). a,b**, Schematic illustrations of OPA chains with alternated bond lengths/densities (**a**, Peierls phase) and uniform bond lengths/densities (**b**, Non-Peierls phase). **c,d**, Schematic illustrations of decaying (**c**, conventional soliton) and non-decaying (**d**, *BFRS*) spectral weight. Insets show the schematics of an OPA chain hybridize (**d**) and does not hybridize (**c**) with the terminals. **e,f**, Total energy as a function of bond length for OPA chains exhibiting BLA (**e**) and those without BLA (**f**).

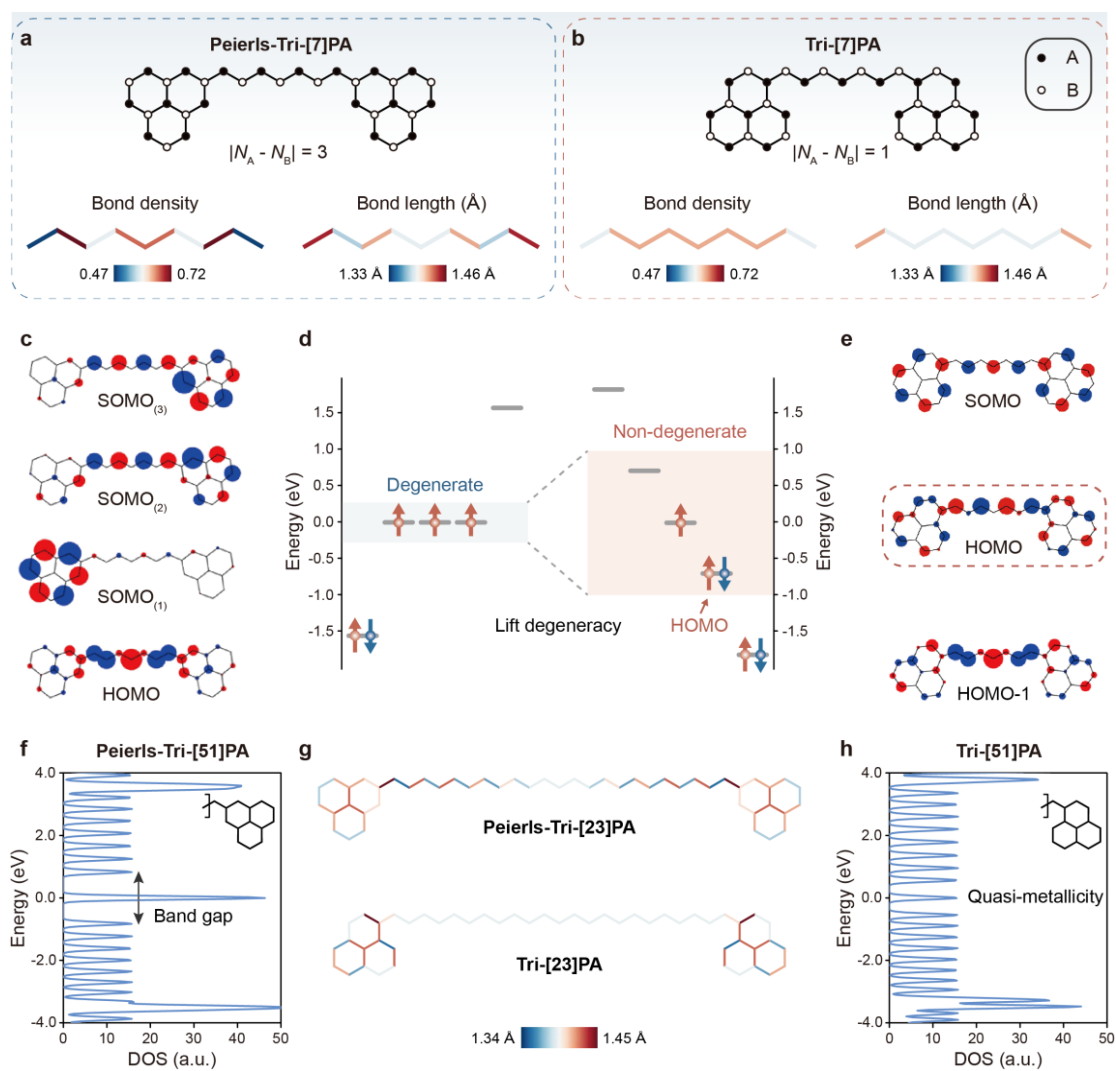

**Fig. 2 | Peierls modeling of OPA chains with different lattice topologies in benzenoid systems. a**,**b**, Structures of **Peierls-Tri-[7]PA** (**a**) and **Tri-[7]PA** (**b**), along with calculated bond densities and bond lengths of their OPA chains using the Peierls model. **c**, Hückel orbitals of HOMO and SOMOs of **Peierls-Tri-[7]PA**. **d**, Hückel spectra of **Peierls-Tri-[7]PA** (left) and **Tri-[7]PA** (right). **e**, Hückel orbitals of HOMO-1, HOMO, and SOMO of **Tri-[7]PA**. **f**,**h**, DFT-calculated DOS around the Fermi level for **Peierls-Tri-[51]PA** (**f**) and **Tri-[51]PA** (**h**). **g**, DFT-calculated bond lengths for **Peierls-Tri-[23]PA** and **Tri-[23]PA**.

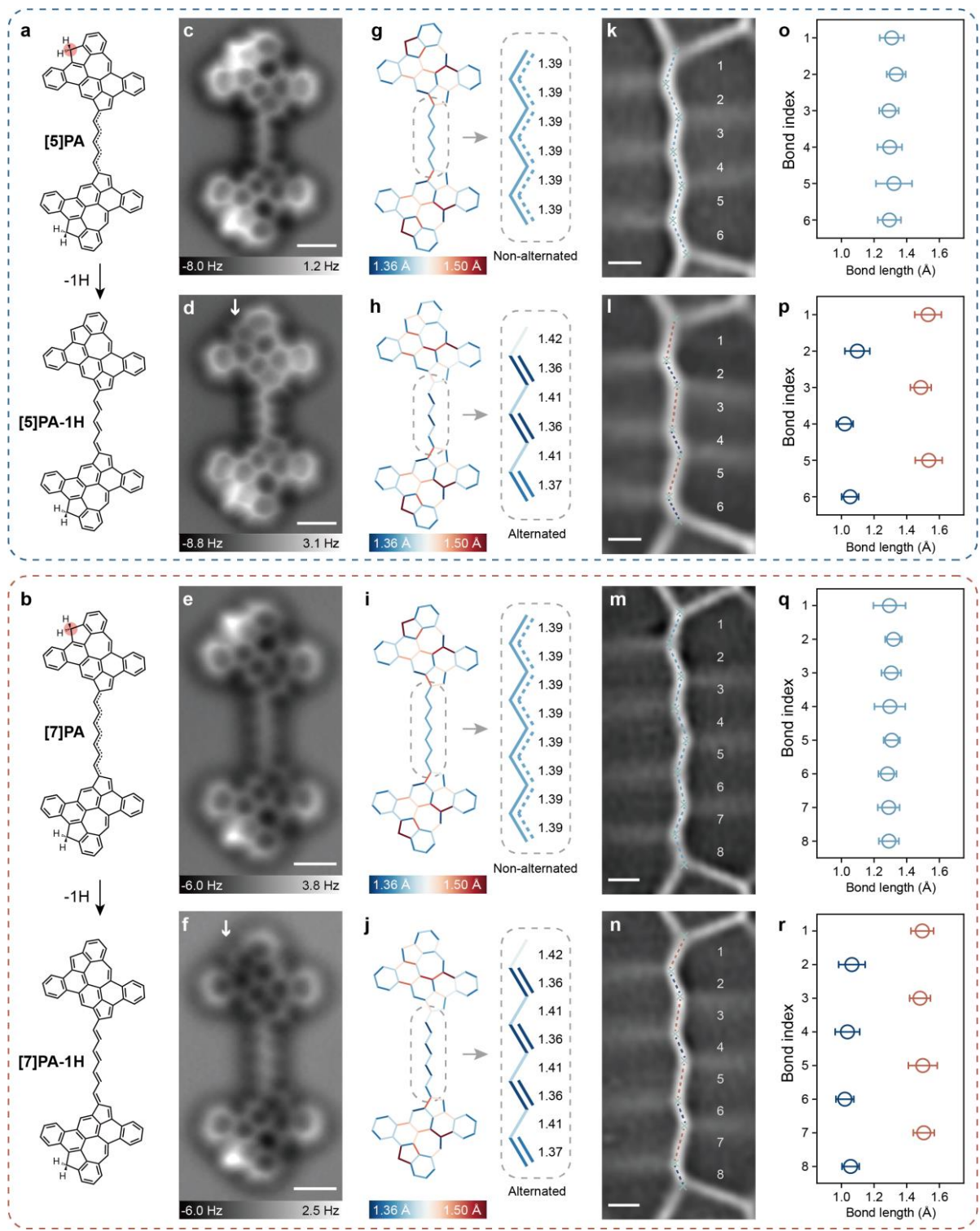

**Fig. 3 | Measuring the bond lengths of C-C bonds in OPA chains. a,b**, Reaction scheme illustrating tip-induced dehydrogenation. **c-f**, Constant-height AFM images of **[5]PA (c)**, **[5]PA-1H (d)**, **[7]PA (e)**, and **[7]PA-1H (f)** ($V$ = 2 mV, $\Delta z$ = -1.6 Å (**c,e**) and -1.8 Å (**d,f**); set

point prior to turning off feedback: $V = 20$ mV, $I = 30$ pA). **g-j**, DFT-calculated bond lengths of **[5]PA (g)**, **[5]PA-1H (h)**, **[7]PA (i)**, and **[7]PA-1H (j)**, plotted in the selected color range, with zoomed-in views of the central OPA chains exhibiting either identical (**g,i**) or alternated (**h,j**) bond lengths. **k-n**, Laplace-filtered AFM images of the central OPA chains in **[5]PA (k)**, **[5]PA-1H (l)**, **[7]PA (m)**, and **[7]PA-1H (n)**, highlighted with dashed lines ($V = 2$ mV, $\Delta z = -2.4$ Å; set point prior to turning off feedback: $V = 20$ mV, $I = 30$ pA). **o-r**, Plots of the measured bond lengths of the indicated bonds in (**k-n**). Scale bars: 0.5 nm (**c-f**) and 0.1 nm (**k-n**).

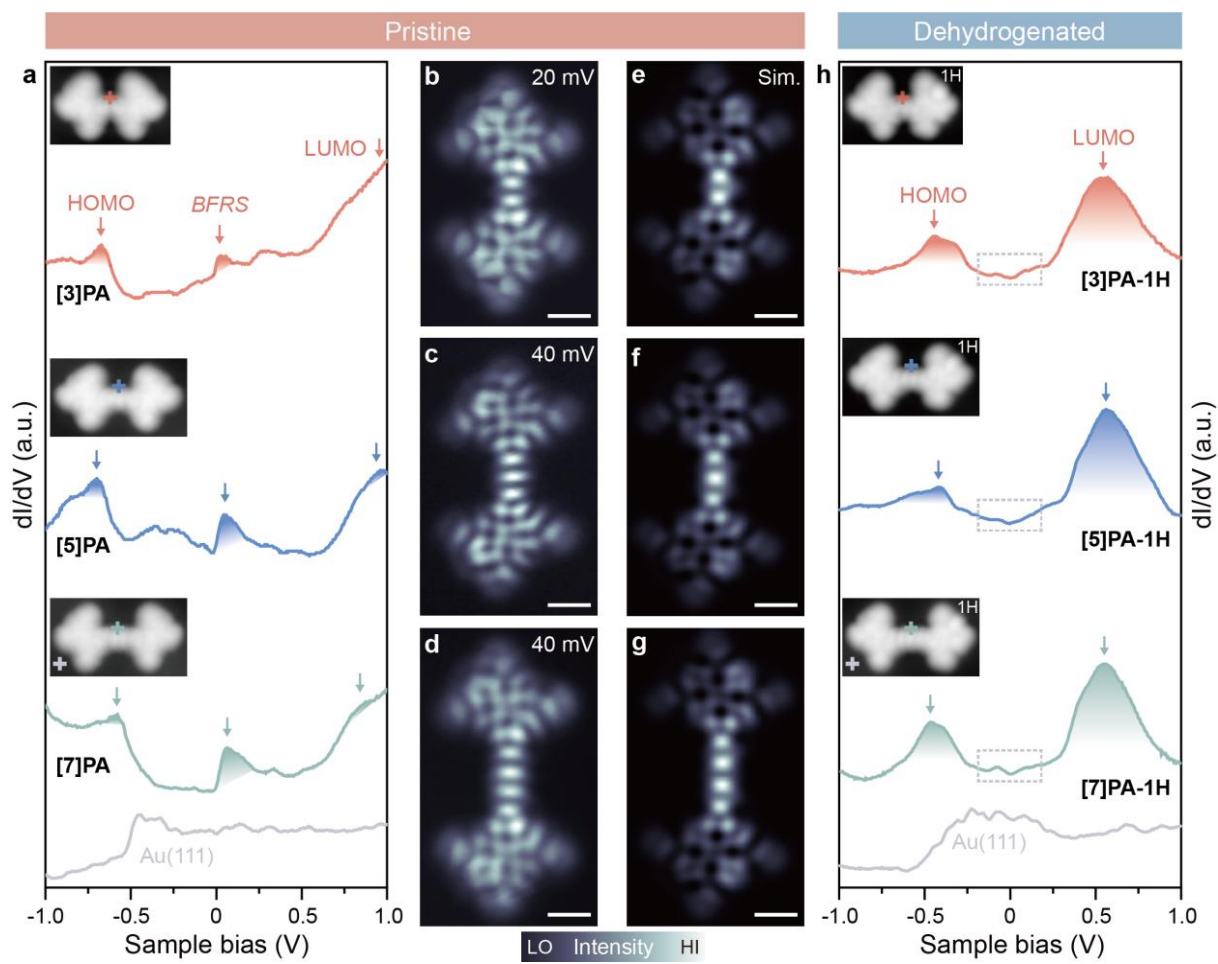

**Fig. 4 | Electronic structures of OPA chains. a**, Point d$I$/d$V$ spectra acquired over the midsection of **[3]PA** (red curve), **[5]PA** (blue curve), **[7]PA** (green curve), and on Au(111) substrate (grey curve). Insets show the corresponding STM images ($V$ = -0.7 V, $I$ = 100 pA). **b-d**, Constant-height d$I$/d$V$ maps showing the *BFRS* of **[3]PA** (**b**, $V$ = 20 meV), **[5]PA** (**c**, $V$ = 40 meV), and **[7]PA** (**d**, $V$ = 40 meV), with $V_{rms}$ = 2 mV. **e-g**, Corresponding DFT-simulated d$I$/d$V$ maps at the energy positions of the *BFRS* (tip-sample height = 5 Å). **h**, Point d$I$/d$V$ spectra acquired over the midsection of **[3]PA-1H** (red curve), **[5]PA-1H** (blue curve), **[7]PA-1H** (green curve), and on Au(111) substrate (grey curve). Insets show the corresponding STM images ($V$ = -0.5 V, $I$ = 100 pA). Scale bars: 0.5 nm.

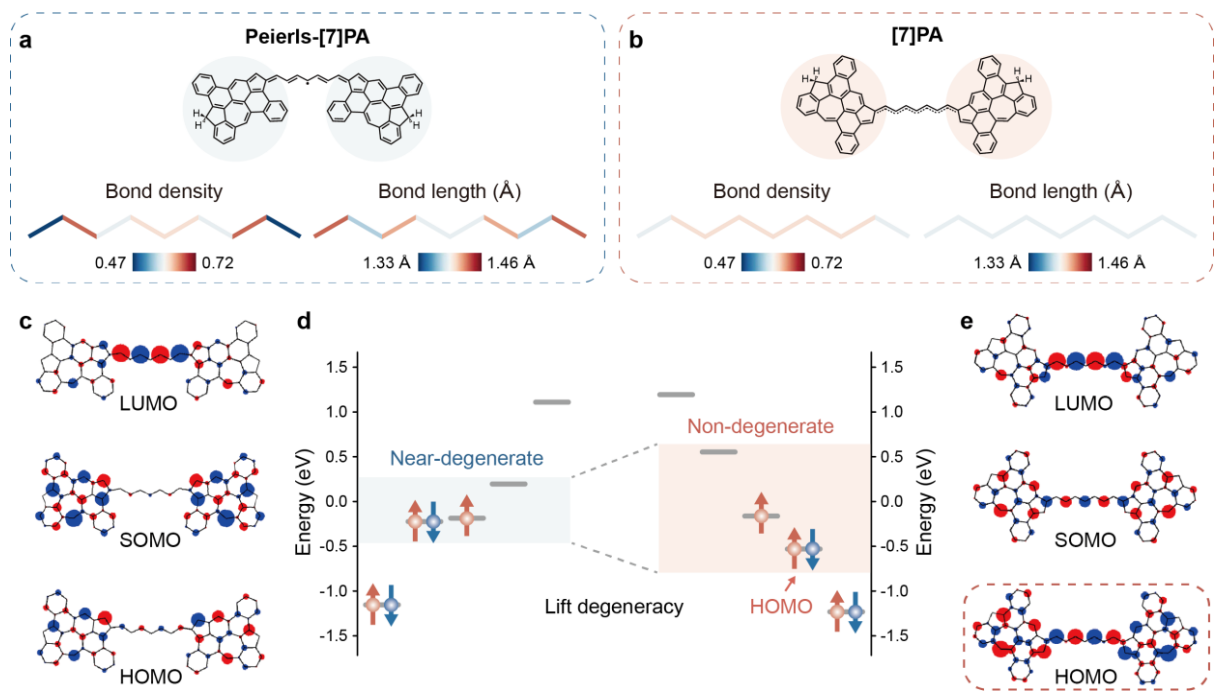

**Fig. 5 | Peierls modeling of OPA chains with different end topologies in non-benzenoid systems. a,b,** Structures of **Peierls-[7]PA** (**a**) and **[7]PA** (**b**), along with calculated bond densities and bond lengths of their OPA chains using the Peierls model. **c**, Hückel orbitals of HOMO, SOMO, and LUMO of **Peierls-[7]PA**. **e**, Hückel spectra of **Peierls-[7]PA** (left) and **[7]PA** (right). **f**, Hückel orbitals of HOMO, SOMO, and LUMO of **[7]PA**.

**Methods**

**Synthesis and sample preparation:**

The detailed synthetic procedures of synthesizing **Precursor** were presented in our previous publication[39]. Au(111) single crystal (MaTeck GmbH) was cleaned by multiple cycles of Ar$^+$ sputtering ($1 \times 10^{-5}$ mbar) and annealing (710 K, 10 min). The precursor was deposited from a Knudsen cell (MBE-Komponenten GmbH) at 470 K onto a clean Au(111) surface held at room temperature. After the deposition of the precursor molecules, the sample was annealed at 500 K for 20 min for the fabrication of **[3]PA**, **[5]PA**, and **[7]PA**. Subsequently, the sample was transferred into the STM head held at 4.4 K for STM/NC-AFM imaging and characterization.

**Experimental details for STM and NC-AFM measurements:**

The STM/NC-AFM experiments were conducted under UHV conditions (base pressure, $<1 \times 10^{-10}$ mbar) at 4.4 K using a Scienta Omicron LT-STM system. We used a qPlus sensor with a resonance frequency $f_0 = 39.65$ kHz, a quality factor $Q \approx 73,000$, and a spring constant $k \approx 1,800$ N m$^{-1}$ operated in frequency-modulation mode[48,49]. The tip apex was functionalized with a CO molecule by picking up CO from the Au(111) surface[50]. The bias voltage ($V$) was applied to the sample with respect to the tip. AFM images were acquired in constant-height mode with $V = 0$ V and an oscillation amplitude ($A$) of 50 pm. The tip-height offsets, $\Delta z$, for constant-height AFM images are defined as the offset in tip-sample distance relative to the STM set point on the Au(111) surface. Positive (negative) values of $\Delta z$ correspond to the tip-sample distance increased (decreased) with respect to the STM set point. The d$I$/d$V$ spectra were collected using an internal lock-in amplifier with a modulation frequency of 879 Hz and an amplitude of 20 mV (unless otherwise noted).

**DFT calculations and simulations:**

We have employed Density Functional Theory (DFT) with the packet FHI-AIMS[51] to relax the systems considered above, restricting z coordinate in order to reproduce the adsorbed planar geometry on Au(111) surface. Relaxations were performed at hybrid GGA-PBE0 level of theory[52] to account for a good description of bond lengths. The convergence settings for forces were set to $10^{-4} eV/\text{Å}$, $10^{-6} eV$ for the total energy, $10^{-4}$ for the charge density and $10^{-3}$ for the sum of Kohn Sham eigenvalues. Scalar relativistic corrections to the kinetic energy of all atoms at the zeroth-order regular approximation (ZORA)[53] were also considered.

AFM simulations were performed by means of the Probe Particle Model[42] with a stiffness of $0.25\ N \cdot m^{-1}$ for the CO tip-probe and an effective charge of $-0.2\ e$. The tip-sample electrostatic interactions were included by the Hartree potential of the system computed by DFT.

For the theoretical d$I$/d$V$ maps we used the Probe Particle Scanning Probe Microscopy model[43]. The CO tip was represented by a linear combination of 10% of s-type orbital and 90% of px and py orbitals in the plane parallel to the sample.

**Data availability**

All data supporting the findings of this study are available within the paper and its online Extended Data file.

**Acknowledgments**

J.L. acknowledges the support from the NRF, Prime Minister's Office, Singapore, under the Competitive Research Program Award (NRF-CRP29-2022-0004), MOE grants (MOE T2EP50121-0008 and MOE-T2EP10221-0005) and Agency for Science, Technology and Research (A*STAR) under its AME IRG Grant (Project715 number M21K2c0113). P.J. acknowledges the support from GACR 20-13692C and the CzechNanoLab Research Infrastructure supported by MEYS CR (LM2023051). J.S. acknowledges the support from the start-up fund (Project ID: P0056699, Number: 1.11.56.WZCK) provided by The Hong Kong Polytechnic University. J.G. acknowledges the support from the National Research Foundation, Singapore and A*STAR under its CQT Bridging. S.S. acknowledges the support from A*STAR under its AME YIRG Grant (M22K3c0094).

**Author contributions**

J.Lu supervised the project and organized the collaboration. X.P. and J.Lu conceived and designed the experiments. X.P. and J.S. carried out the STM/NC-AFM measurements with the help from S.S. J.W. and J.L. synthesized the organic precursors. P.J. conceived and supervised the theoretical analyses and calculations. M.L., L.W., D.S.P. and T.T. performed the theoretical studies and DFT calculations. F.J.G. assisted in the optimization of the NC-AFM sensor. X.P., M.L., P.J. and J. Lu wrote the manuscript with inputs from all authors.

**Competing interests**

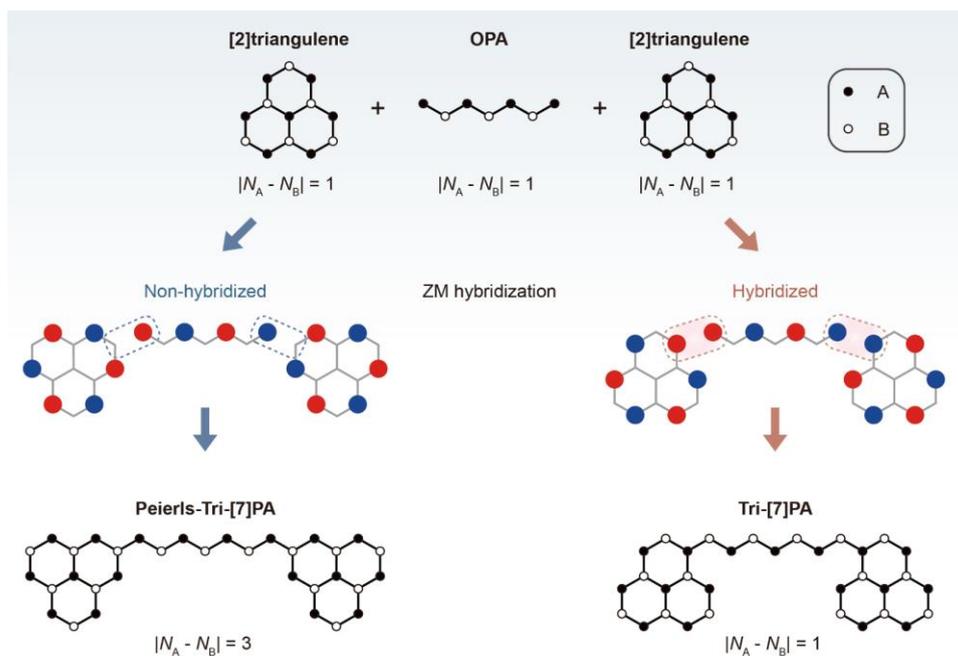

**Extended Data Fig. 1 | Schematic illustration of possible end topologies of an OPA chain terminated with [2]triangulene.** Illustration of effective hybridization between the ZMs of [2]triangulene and the ZM of OPA chain in **Tri-[7]PA** (right), compared to ineffective hybridization in **Peierls-Tri-[7]PA** (left).

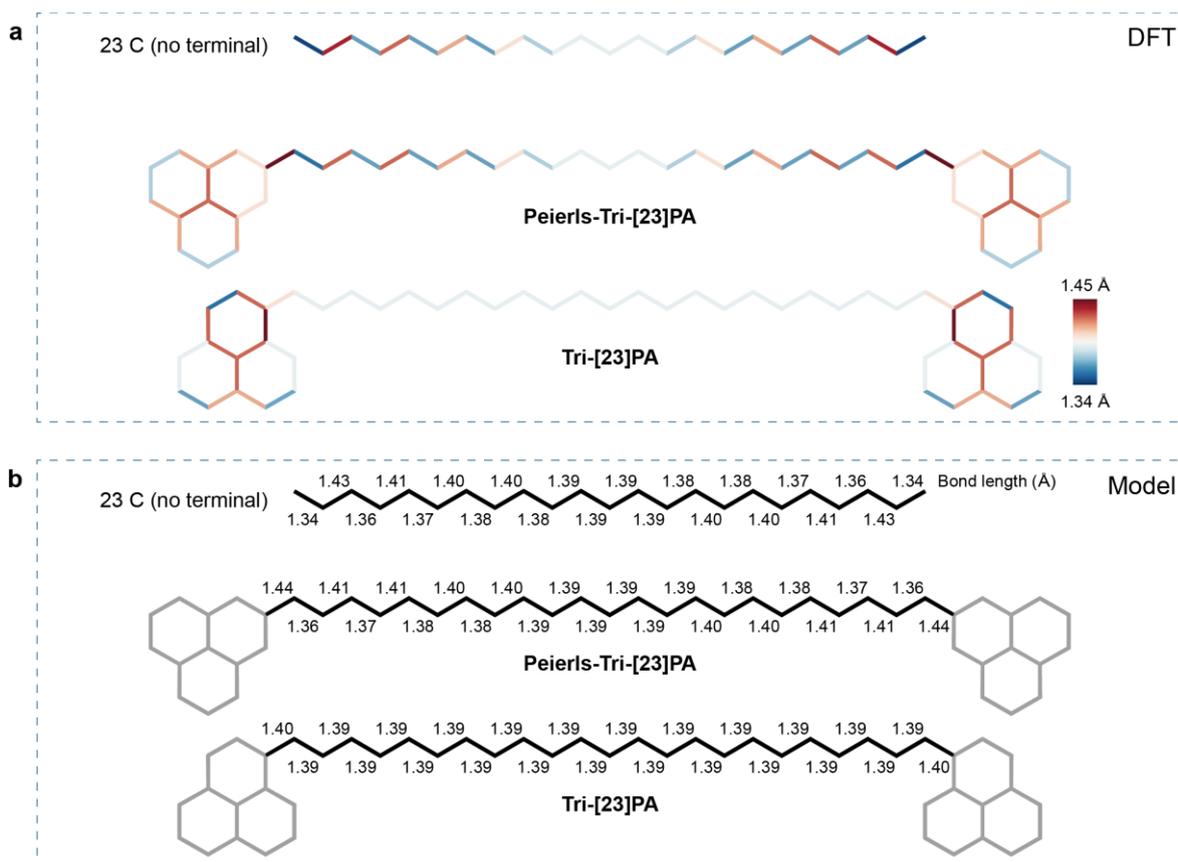

**Extended Data Fig. 2 | Calculated bond lengths for long OPA chains with [2]triangulene terminals featuring different lattice topologies. a**, DFT-calculated bond lengths for the OPA chains consisting of 23 carbon atoms with no terminal (top), with lattice topology matching that of **Peierls-Tri-[7]PA** (middle, referred to as **Peierls-Tri-[23]PA**), and with lattice topology matching that of **Tri-[7]PA** (bottom, referred to as **Tri-[23]PA**) plotted in the selected color range. All structures are completely relaxed by total-energy DFT calculations. **b**, Corresponding calculated bond lengths by the Peierls model described, showing agreement with DFT results.

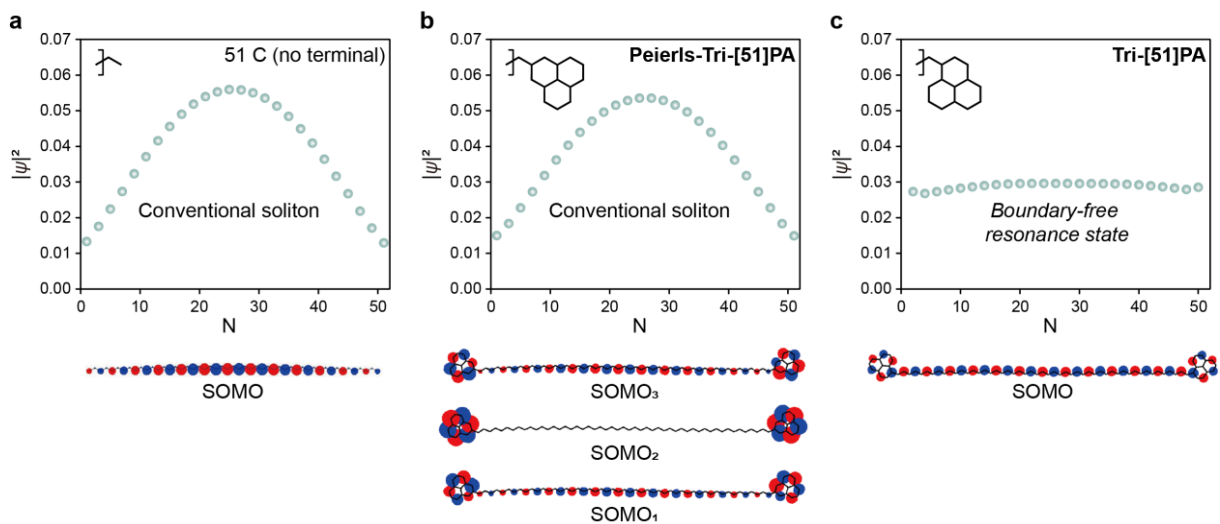

**Extended Data Fig. 3 | Distributions of in-gap states in OPA chains with different lattice topologies.** Plotted squared wavefunction at the carbon atoms of OPA chains consisting of 51 carbon atoms with no terminal (**a**), **Peierls-Tri-[51]PA** (**b**), and **Tri-[51]PA** (**c**). The squared wavefunction of the degenerate soliton states (SOMO$_{1-3}$) of **Peierls-Tri-[51]PA** are summed. All wavefunctions are normalized on the backbone and localized at the carbon atoms. Bottom panels show the corresponding Hückel orbitals of the soliton states (SOMOs).

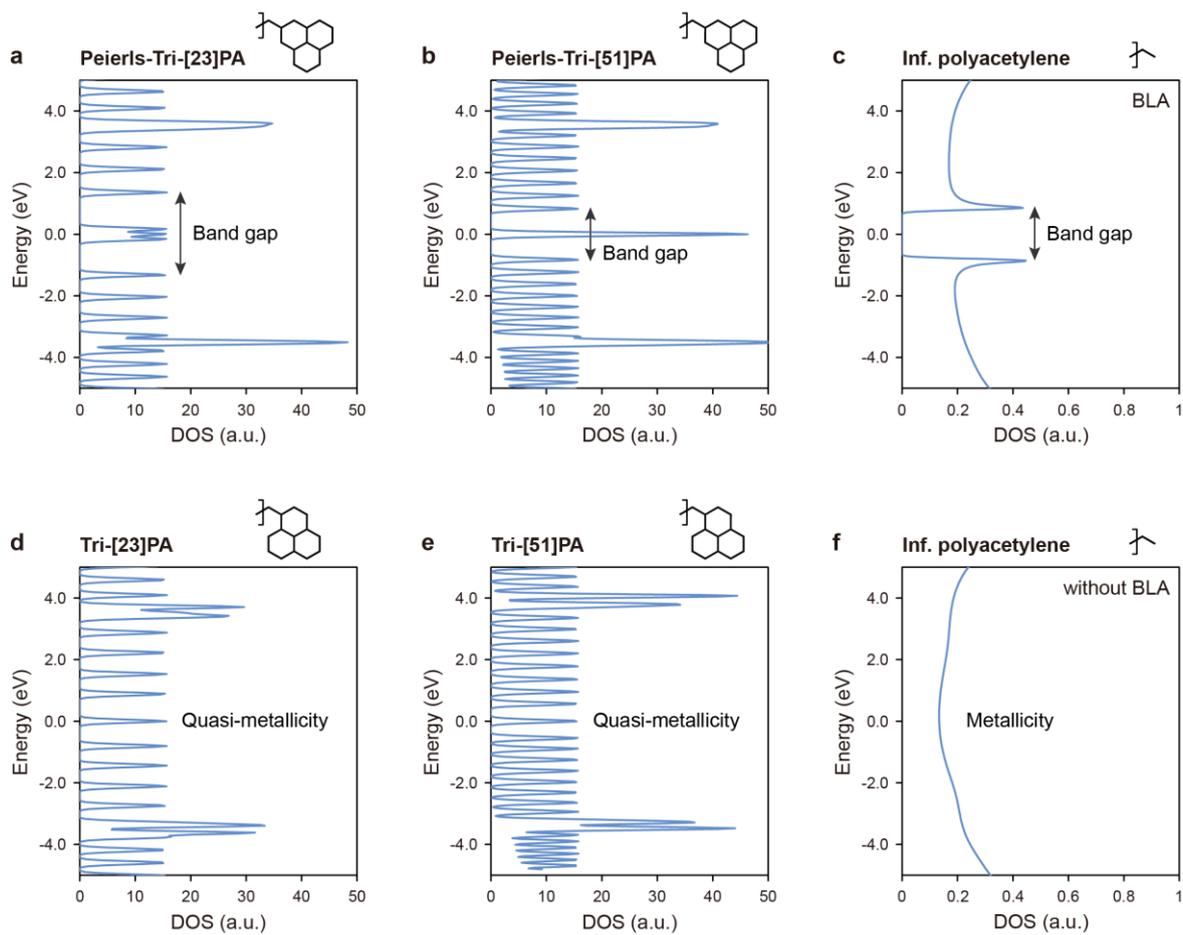

**Extended Data Fig. 4 | Calculated density of states (DOS) near the Fermi level for polyacetylene chains with different lengths and lattice topologies. a-c**, Calculated DOS for **Peierls-Tri-[23]PA** (**a**), **Peierls-Tri-[51]PA** (**b**) with lattice topology matching that of **Peierls-[7]Tri-PA**, and an infinite polyacetylene chain with BLA (**c**). The presence of three ZMs creates an enhanced in-gap peak in the DOS. **d-f**, Calculated DOS for **Tri-[23]PA** (**d**), **Tri-[51]PA** (**e**) with lattice topology matching that of **Tri-[7]PA**, and an infinite polyacetylene chain with equidistant bond length of 1.39 Å (**f**).

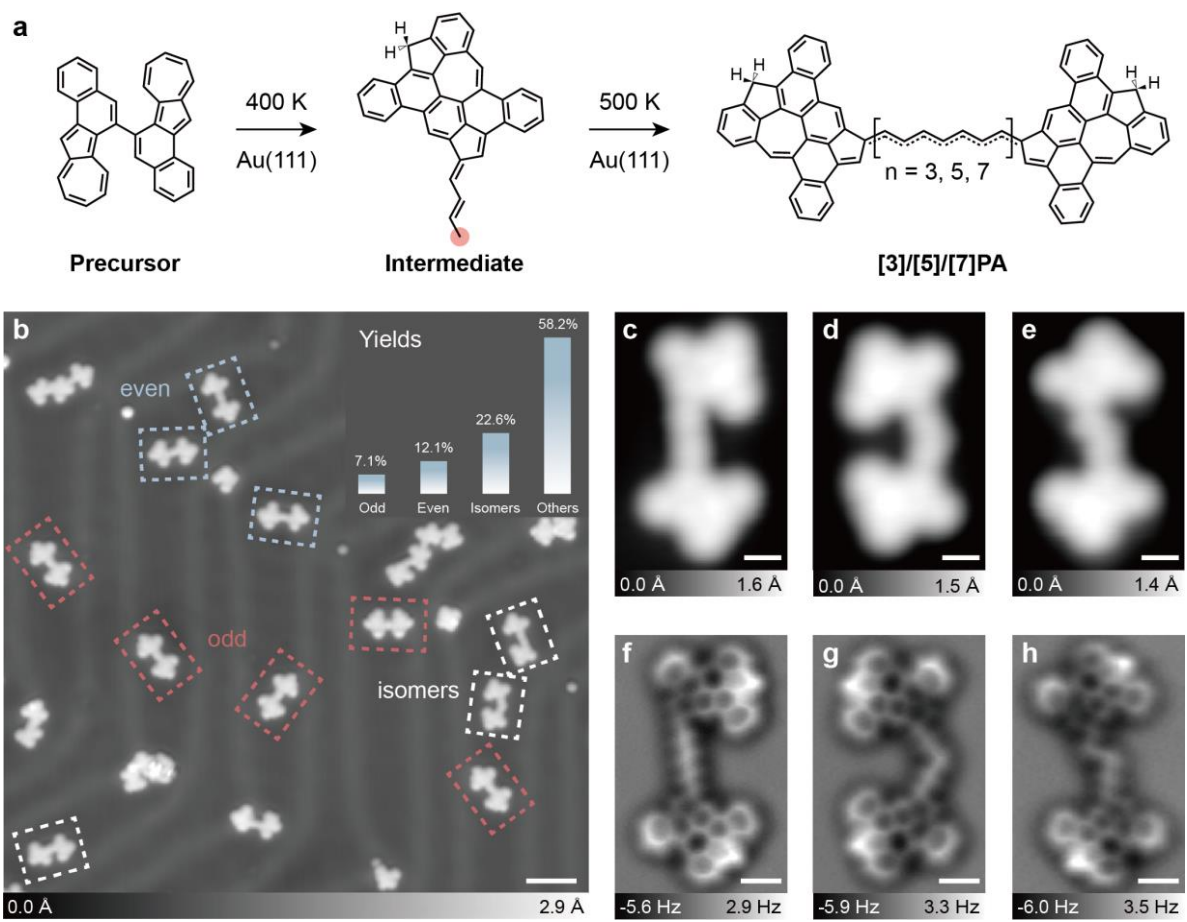

**Extended Data Fig. 5 | On-surface synthesis of OPA chains. a**, Synthetic pathway from **Precursor** to **[3]/[5]/[7]PA** with a reported **Intermediate**[39]. **b**, Overview STM image of products after annealing the precursor on Au(111) at 500 K ($V = 0.1$ V, $I = 100$ pA). Products with odd- and even-membered polyacetylene chains, as well as isomers, are highlighted with red, blue, and white dashed boxes, respectively. Inset bar plots show product yields based on statistics from over 500 molecules. **c-e**, STM images of three distinct isomers arising from trans-cis isomerization of the central OPA chain ($V = 0.1$ V, $I = 100$ pA). **f-h**, Corresponding constant-height AFM images of isomers ($V = 2$ mV, $\Delta z = -1.6$ Å; set point prior to turning off feedback: $V = 20$ mV, $I = 30$ pA). Scale bars: 3 nm (**b**) and 0.5 nm (**c-h**).

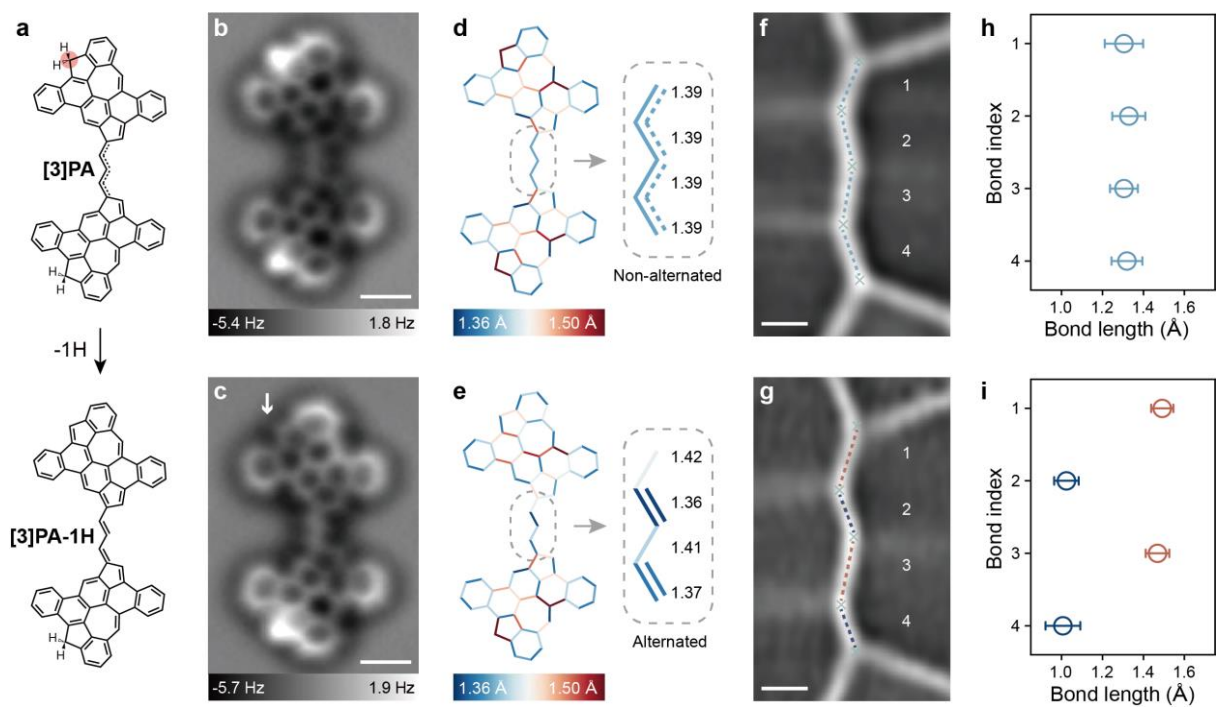

**Extended Data Fig. 6 | Measuring the bond lengths of C-C bonds in OPA chains ([3]PA). a**, Reaction scheme illustrating tip-induced dehydrogenation. **b-c**, Constant-height AFM images of **[3]PA** (**b**) and **[3]PA-1H** (**c**) ($V = 2$ mV, $\Delta z = -1.6$ Å; set point prior to turning off feedback: $V = 20$ mV, $I = 30$ pA). **d-e**, DFT-calculated bond lengths of **[3]PA** (**d**) and **[3]PA-1H** (**e**), plotted in the selected color range, with zoomed-in views of the central OPA chains exhibiting either identical (**d**) or alternated (**e**) bond lengths. **f-g**, Laplace-filtered AFM images of the central OPA chains in **[3]PA** (**f**) and **[3]PA-1H** (**g**), highlighted with dashed lines ($V = 2$ mV, $\Delta z = -2.4$ Å; set point prior to turning off feedback: $V = 20$ mV, $I = 30$ pA). **h-i**, Plots of the measured bond lengths of the indicated bonds in (**f-g**). Scale bars: 0.5 nm (**b-c**) and 0.1 nm (**f-g**).

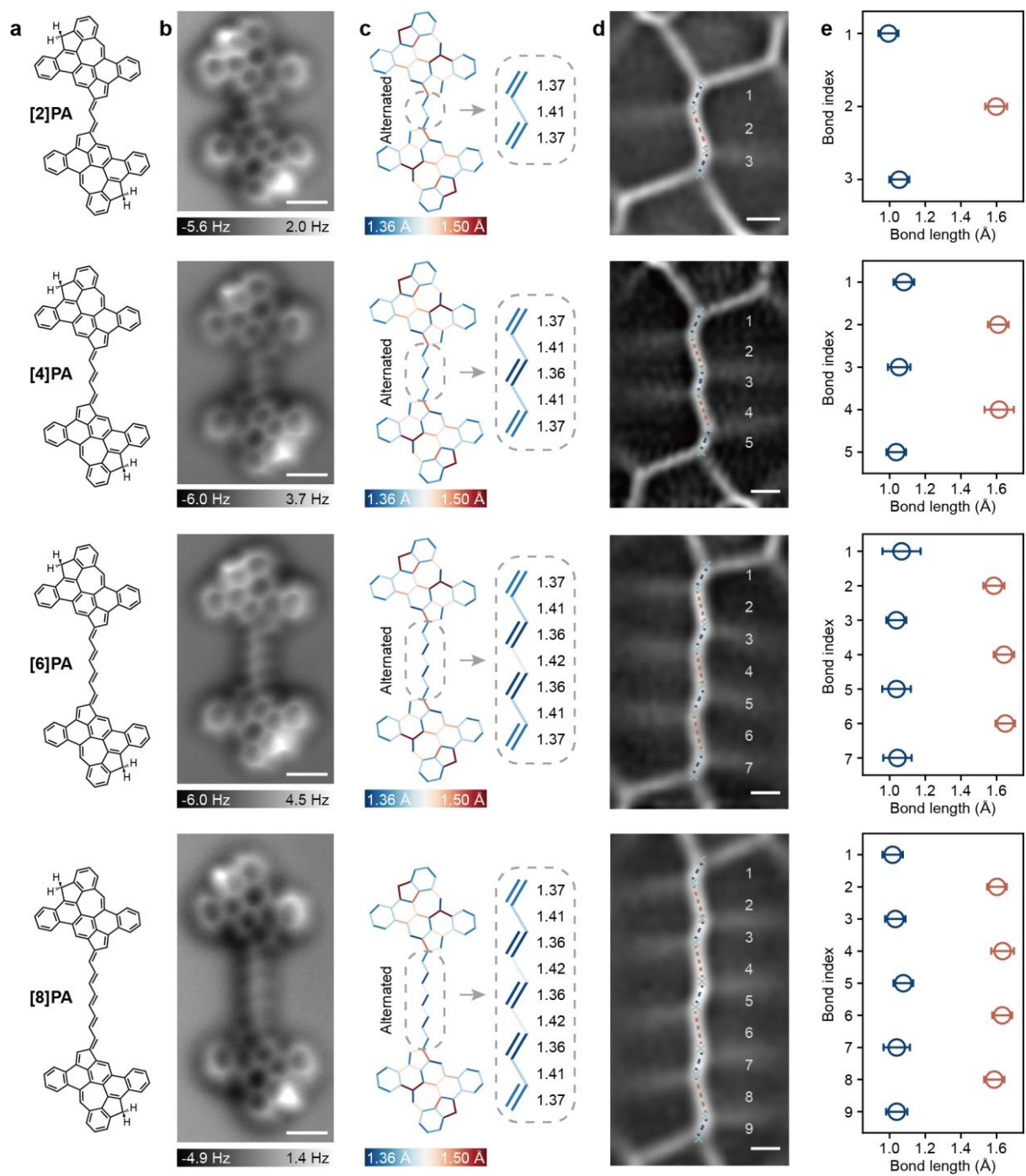

**Extended Data Fig. 7 | Measuring the bond lengths of C-C bonds in EPA chains. a**, Chemical structures of **[2]PA**, **[4]PA**, **[6]PA**, and **[8]PA**, respectively. **b**, Constant-height AFM images of **[2]PA**, **[4]PA**, **[6]PA**, and **[8]PA**, respectively. ($V$ = 2 mV, $\Delta z$ = -1.6 Å; set point prior to turning off feedback: $V$ = 20 mV, $I$ = 30 pA). **c**, DFT-calculated bond lengths of **[2]PA**, **[4]PA**, **[6]PA**, and **[8]PA**, plotted in the selected color range, with zoomed-in views of the central EPA chains exhibiting alternated bond lengths. **d**, Laplace-filtered AFM images

of the central EPA chains in **[2]PA**, **[4]PA**, **[6]PA**, and **[8]PA**, highlighted with dashed lines ($V = 2$ mV, $\Delta z = -2.4$ Å; set point prior to turning off feedback: $V = 20$ mV, $I = 30$ pA). **e**, Plots of the measured bond lengths of the indicated bonds in (**d**). Scale bars: 0.5 nm (**b**) and 0.1 nm (**d**).

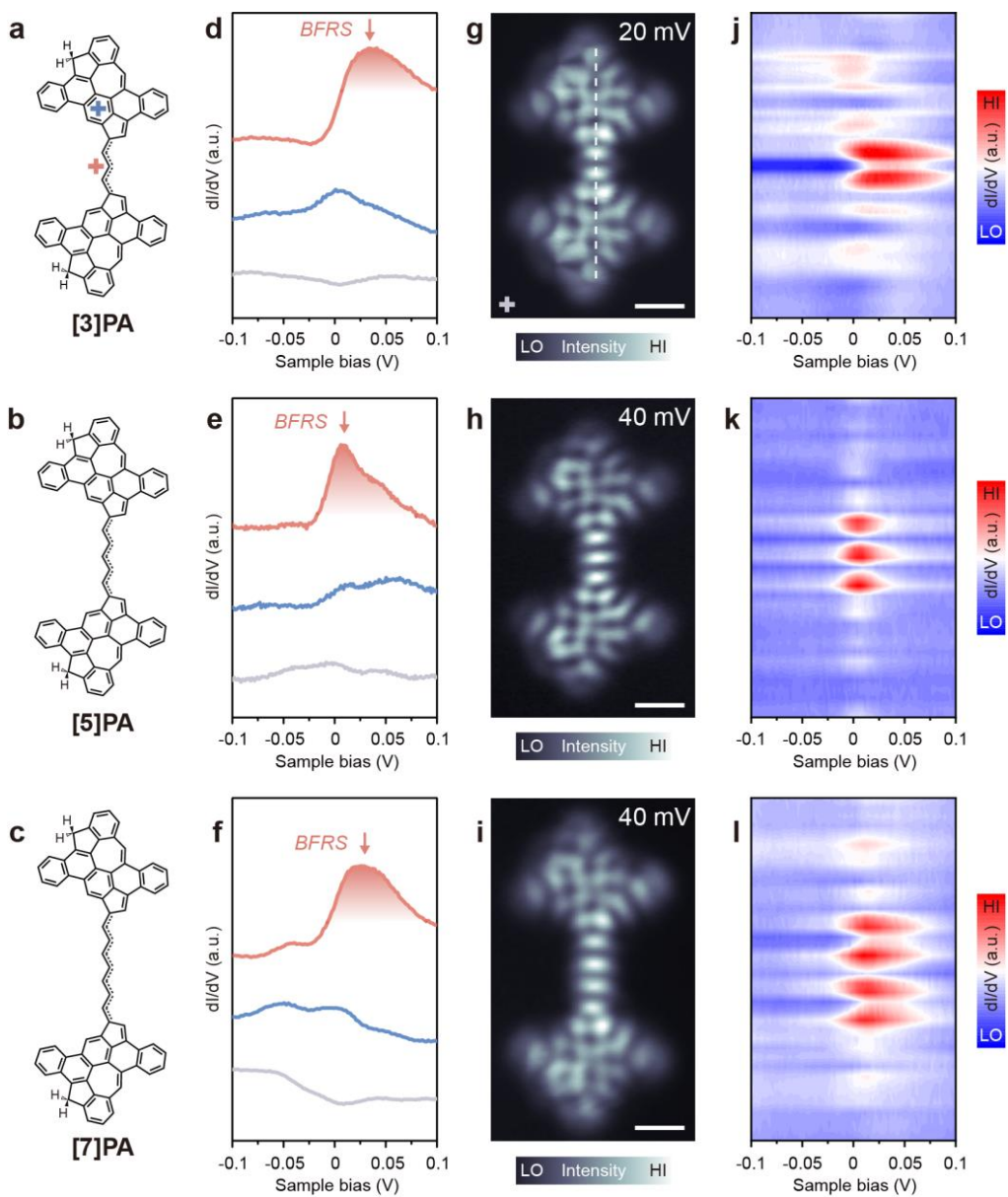

**Extended Data Fig. 8 | Characterization of non-decaying *boundary-free resonance states*.**
**a-c**, Chemical structures of the pristine molecules: **[3]PA** (**a**), **[5]PA** (**b**), and **[7]PA** (**c**). **d-f**, Corresponding point d$I$/d$V$ spectra acquired over the central OPA chains (red curve), nanographene terminals (blue curve), and on Au(111) substrate (grey curve). **g-i**, Corresponding constant-height d$I$/d$V$ maps showing the *BFRS* ($V_{rms}$ = 2 mV). **j-l**, Corresponding colour-coded d$I$/d$V$ spectra acquired across the molecules. The actual positions where the spectra were taken are illustrated by a grey dashed line in the d$I$/d$V$ map in **g**. Scale bars: 0.5 nm.

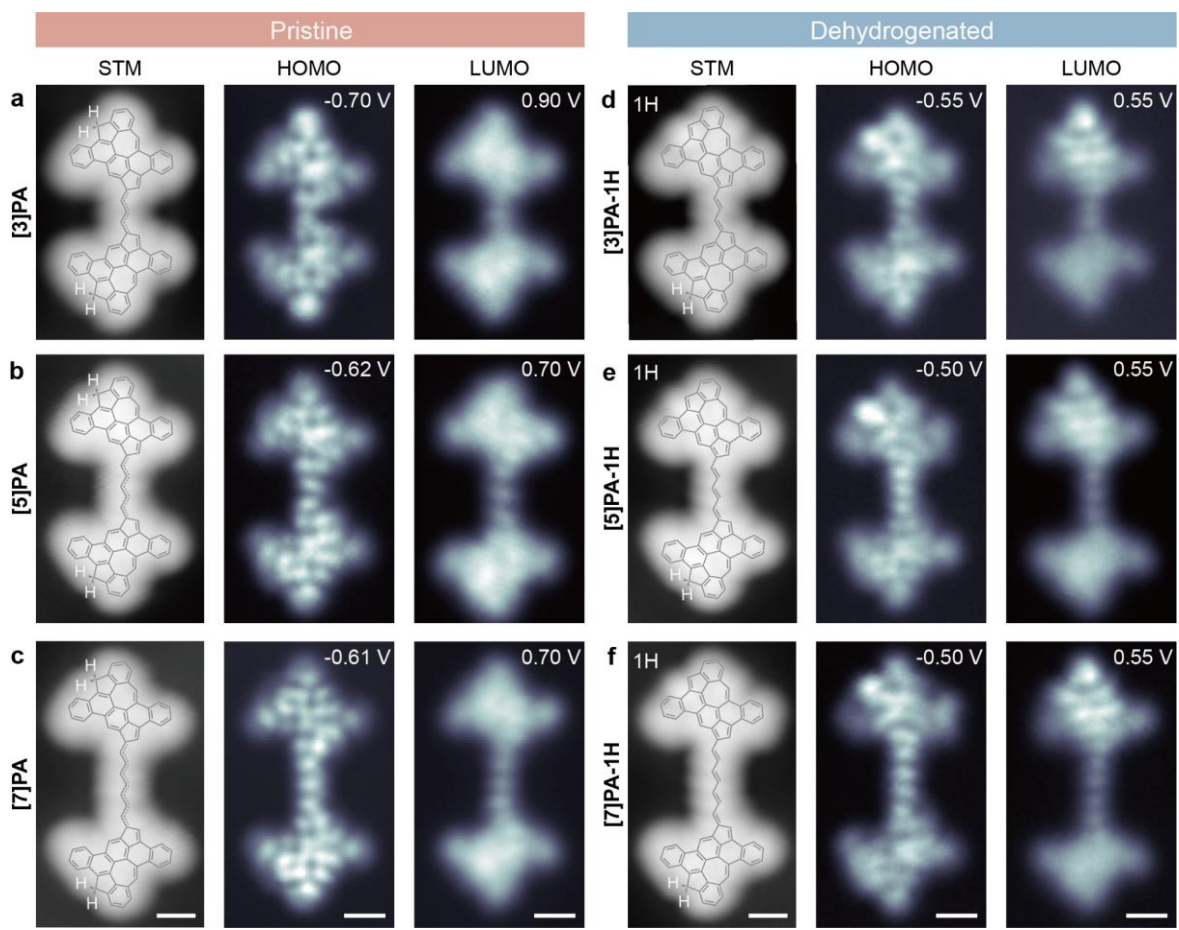

**Extended Data Fig. 9 | d*I*/d*V* maps of OPA chains and their dehydrogenated products. a-f**, Corresponding STM images (left) (*V* = -0.7 V, *I* = 100 pA) and constant-height d*I*/d*V* maps of the HOMO (middle) and LUMO (right) of **[3]PA (a)**, **[5]PA (b)**, **[7]PA (c)**, **[3]PA-1H (d)**, **[5]PA-1H (e)**, and **[7]PA-1H (f)** (*I* = 1 nA, $V_{rms}$ = 20 mV). Scale bars: 0.5 nm.

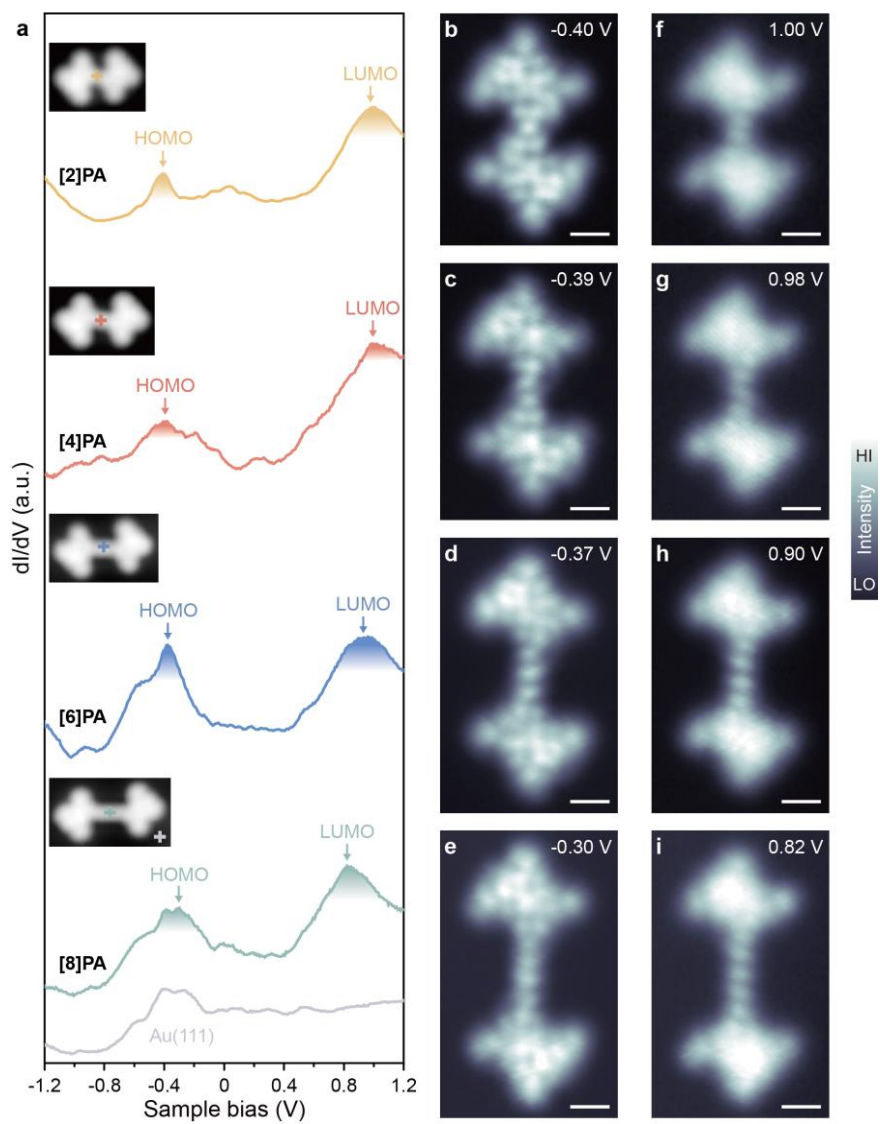

**Extended Data Fig. 10 | Electronic structures of EPA chains. a**, Point d$I$/d$V$ spectra acquired over the midsection of **[2]PA** (yellow curve), **[4]PA** (red curve), **[6]PA** (blue curve), **[8]PA** (green curve), and on Au(111) substrate (grey curve). Insets show the corresponding STM images ($V$ = 0.1 V, $I$ = 100 pA). **b-e**, Constant-height d$I$/d$V$ maps of the HOMO for **[2]PA (b)**, **[4]PA (c)**, **[6]PA (d)**, and **[8]PA (e)** ($I$ = 1 nA, $V_{rms}$ = 20 mV). **f-i**, Constant-height d$I$/d$V$ maps of the LUMO for **[2]PA (f)**, **[4]PA (g)**, **[6]PA (h)**, and **[8]PA (i)** ($I$ = 1 nA, $V_{rms}$ = 20 mV). Scale bars: 0.5 nm.